\documentstyle[11pt,aaspp4]{article}
\slugcomment{Accepted by the Astrophysical Journal}

\lefthead{G\'{o}mez et al.}
\righthead{A665}

\def\rosat{{\it ROSAT}}
\def\chandra{{\it Chandra}}
\def\xmm{{\it XMM}}

\def\asca{{\it ASCA}}
\def\lsim{\hbox{\raise.35ex\rlap{$<$}\lower.6ex\hbox{$\sim$}\ }}

\begin{document}

\title{A Merger Scenario for the Dynamics of Abell
665\footnote{Optical observations reported here were
  obtained at the Multiple Mirror Telescope, a joint facility of the
  Smithsonian Institution and the University of Arizona.}}

\author{Percy L. G\'{o}mez, John P. Hughes\altaffilmark{2}}
\affil{Department of Physics and Astronomy, Rutgers The State
 University of New Jersey, \\
 136 Frelinghuysen Road, Piscataway NJ 08854-8019\\  
 E-mail: percy@physics.rutgers.edu, jph@physics.rutgers.edu}
\altaffiltext{2}{
Also Service d'Astrophysique, L'Orme des Merisiers, 
CEA-Saclay, 91191 Gif-sur-Yvette Cedex France
}

\author{Mark Birkinshaw}
\affil{Department of Physics, University of Bristol, Bristol, BS8 1TL, UK\\
     E-mail: mark.birkinshaw@bristol.ac.uk}

% Notice that each of these authors has alternate affiliations, which
% are identified by the \altaffilmark after each name.  The actual alternate
% affiliation information is typeset in footnotes at the bottom of the
% first page, and the text itself is specified in \altaffiltext commands.
% There is a separate \altaffiltext for each alternate affiliation
% indicated above.

% The abstract environment prints out the receipt and acceptance dates
% if they are relevant for the journal style.  For the aasms style, they
% will print out as horizontal rules for the editorial staff to type
% on, so long as the author does not include \received and \accepted
% commands.  This should not be done, since \received and \accepted dates
% are not known to the author.

\begin{abstract}

We present new redshift measurements for 55 galaxies in the vicinity
of the rich galaxy cluster Abell 665. When combined with results from
the literature, we have good velocity measurements for a sample of 77
confirmed cluster members from which we derive the cluster's redshift
$z=0.1829\pm 0.0005$ and line-of-sight velocity dispersion $\sigma =
1390^{+120}_{-110}\,\rm km\, s^{-1}$. Our analysis of the kinematical
and spatial data for the subset of galaxies located within the central
750 kpc reveals only subtle evidence for substructure and
non-Gaussianity in the velocity distribution. We find that the
brightest cluster member is not moving significantly relative to the
other galaxies near the center of the cluster.  On the other hand, our
deep \rosat\ high resolution image of A665 shows strong evidence for
isophotal twisting and centroid variation, thereby confirming previous
suggestions of significant substructure in the hot X-ray--emitting
intracluster gas. In light of this evident substructure, we have
compared the optical velocity data with N-body simulations of head-on
cluster mergers. We find that a merger of two similar mass subclusters
(mass ratios of 1:1 or 1:2) seen close to the time of core-crossing
produces velocity distributions that are consistent with that observed.

\end{abstract}

% The different journals have different requirements for keywords.  The
% keywords.apj file, found on aas.org in the pubs/aastex-misc directory, 
% contains a list of keywords used with the ApJ and Letters.  These are 
% usually assigned by the editor, but authors may include them in their 
% manuscripts if they wish. 


% That's it for the front matter.  On to the main body of the paper.
% We'll only put in tutorial remarks at the beginning of each section
% so you can see entire sections together.

% In the first two sections, you should notice the use of the LaTeX \cite
% command to identify citations.  The citations are tied to the
% reference list via symbolic KEYs.  We have chosen the first three
% characters of the first author's name plus the last two numeral of the
% year of publication.  The corresponding reference has a \bibitem
% command in the reference list below.
%
% Please see the AASTeX manual for a more complete discussion on how to make
% \cite-\bibitem work for you.   

\section{INTRODUCTION}

The study of galaxy clusters has revealed them as powerful probes of
such cosmological quantities as the baryon fraction of the Universe
(e.g., White \& Fabian 1995), $\Omega_0$ (Richstone et al.~1992), and
the Hubble constant $H_0$ (Gunn 1978; Birkinshaw 1979).  However, our
knowledge of the physics of galaxy clusters has not yet reached the
same level of understanding that we have gained for, arguably, the
most important cosmological probe to date, i.e., Cepheid variables,
and this remains as one of the most significant limitations in our use
of clusters for cosmological studies. For instance, one can compute
$H_0$ by combining measurements of the decrement in the brightness
temperature of the cosmic microwave background radiation (CMBR) caused
by the inverse Compton scattering of CMBR photons by the hot electrons
in the cluster (the Sunyaev-Zel'dovich effect, Sunyaev \& Zel'dovich
1972) with spectral and imaging observations of the X-ray emission
produced by the same hot gas. This technique requires accurate 3-D
models of the properties (temperature, density, metallicity, etc.) of
the cluster atmosphere, which in turn demands that the physics and
astrophysics of clusters be well understood.  However, this
understanding presents a problem because even present-day
clusters are dynamically young and active, showing evidence for the
accretion and merger of other systems. The resulting rich
complexity in their internal properties greatly complicates their
use for precision cosmology.

Abell 665 was initially classified by Abell (1958) as the richest
cluster in his catalog. As such, it has been the subject of
considerable study across the wavebands. Evidence for subclustering in
the spatial distribution of the galaxies was first presented by Geller
\& Beers (1982). This was later confirmed through BVR photometry of
178 galaxies by Kalloglyan et al.~(1990). The luminosity function has
been recently studied by Garilli et al.~(1996), Wilson et al.~(1997),
and Trentham (1998) for comparison with other clusters and in an
effort to evaluate models for galaxy evolution.

Despite its richness, only 33 cluster members in A665 have published
redshifts (Oegerle et al.~1991, hereafter OHFH). OHFH examined the
kinematical properties of these galaxies and found that the velocity
distribution was well described by a Gaussian with a
relativistically-corrected line-of-sight velocity dispersion of
$\sigma = 1201_{-126}^{+183}\, \rm km\, s^{-1}$.  This seemed to
indicate a fairly relaxed, but massive cluster. However, these authors
did not completely reject the possibility that the cluster could be
more complex.  They suggested that the observed spatial substructure
in the galaxy distribution and the relatively large peculiar velocity
(v$_{pec}$ = 447 km s$^{-1}$) of the brightest cluster member (BCM)
argued for a non-relaxed dynamical state for A665. We will have more
to say on this point later.

A665 is also a luminous source of X-ray emission, as its optical
richness and high velocity dispersion would suggest.  The presence of
hot gas in this cluster was demonstrated by early observations
performed by the {\it Einstein Observatory} and the {\it Ginga}
satellite (Birkinshaw, Hughes, \& Arnaud 1991; Hughes \& Tanaka 1992).
These pioneering observations revealed some evidence that the spatial
distribution of the X-ray--emitting gas deviated from circular
symmetry. Birkinshaw et al.~modeled the complex gas distribution and
found that a combination of two isothermal-$\beta$ models separated by
$\sim$3$^\prime$ provided a considerably better fit to the data than a
single isothermal-$\beta$ model did. Hughes \& Birkinshaw (1996, 2000)
have applied this type of modeling in much greater detail to \rosat\
PSPC data and came to the conclusion that the properties of the X-ray
emission required that a major merger was occurring or had occurred
recently in this system. Note that Markevitch (1996) also suggested
that a recent merger could explain the temperature gradient and
asymmetric X-ray emission detected in \asca\ observations of A665.

Further support for a merger scenario comes from Buote \& Tsai's
(1996) study of the X-ray morphologies of a sample of 59 bright X-ray
clusters. Their work quantifies cluster substructure by characterizing
the X-ray surface brightness in terms of a multipole expansion of the
two-dimensional gravitational potential.  They detect a tight
correlation between specific multipole power ratios that suggests an
evolutionary track for clusters (i.e., the location of a cluster in
the $P_2/P_0$--$P_4/P_0$ power ratio plane is a function of its
dynamical state). Interestingly, Buote \& Tsai find that the only
cluster in their entire sample that deviates from this correlation is
A665. They hypothesized that A665 is undergoing a major merger event
and is in a brief period of its evolution when the X-ray emitting hot
gas does not follow the dark matter distribution or the gravitational
potential.

The goal of this paper is to present new velocity measurements for
galaxies in A665 and, by combining them with data from the literature, to study the cluster's 
dynamics and to investigate the parameters of
the cluster merger that we infer to have taken place recently. In
section~2 we describe the acquisition and reduction of new optical and
\rosat\ X-ray data. In the following section we present our analysis
of these data. Then in section 4 we present our model for the
dynamical state of A665. We summarize our conclusions in section 5. We
use $H_0$=75 km s$^{-1}$ Mpc$^{-1}$ and q$_0$=0.5 throughout the paper. 

\section{OBSERVATIONS AND DATA REDUCTION}

\subsection{Optical}

Initially we used a 0.6 m telescope CCD image to identify and measure
the positions of a total of 147 objects in the inner 5\farcm5 region
of A665 to a relative position uncertainty of 0\farcs5. Later we
obtained rough R band photometry for these objects from images
obtained on 31 December 1991 and 6 March 1992 at the Fred L.~Whipple
Observatory 1.2 m telescope on Mt.~Hopkins. At that time, the
telescope was equipped with a thick, front-illuminated Loral
2048$\times$2048 CCD with a nominal pixel size of
$\sim$0$\farcs$32. Its $\sim$10$\arcmin$ field-of-view allowed us to
mosaic a large area of the cluster by observing the cluster center and
four regions offset by $\sim 5\arcmin$ to the NW, NE, SW, and SE. Each
set of observations consisted of six individual 300 s exposures (with
small positional offsets between frames) in order to reduce cosmic-ray
contamination and cosmetic defects in the CCD chips.  Each image was
bias-subtracted and flat-fielded using a dome flat and then a sky flat
was constructed from the median combination of all individual frames.

A grayscale plot of the final image is shown in Figure 1. From this image 
we computed R band magnitudes for the galaxy targets within a fixed
5$\arcsec$ radius aperture. Short observations of Landolt (1992)
standard stars allowed us to reduce our measurements to the
Johnson-Kron-Cousins system.  To determine the zero point offset we
compared our measurements with the magnitudes from a photometric,
calibrated sample of galaxies in A665 (Trentham 1998). We find a
root-mean-square (rms) dispersion of $m_{\rm R} \sim 0.13$ in the
magnitudes of the 55 galaxies in common between our and Trentham's
sample. We consider this to be the uncertainty in the overall accuracy
of our photometry.

Of the 147 potential targets in our sample, we observed a total of 89
objects during seven spectroscopic observing runs at the Multiple
Mirror Telescope (MMT). Table 1 lists the combination of
spectrographs, gratings, and aperture masks used in these runs. The
atmospheric conditions were generally good, except for the March 1990
observing session, which was hampered by bad weather and poor seeing.
The first three runs used the faint-object grism spectrograph (FOGS)
(Geary, Huchra, \& Latham 1986) while the remaining spectra were
acquired with the Red Channel spectrograph (Schmidt, Weymann, \& Foltz
1989).

Two gratings, one with 400 lines mm$^{-1}$ and the other one with 300
lines mm$^{-1}$, were used with the FOGS. They offered resolutions of
11\AA\ and 15\AA\ over nominal spectral ranges of 4000\AA--6700\AA\
and 3800\AA--7500\AA, respectively. The red channel spectrograph was
equipped with a 270 lines mm$^{-1}$ grating that provided a typical
spectral resolution of $\sim$11\AA\ over 3800\AA--7400\AA\
(nominal). Note that the exact spectral range depended on the location
in the focal plane of the slit for each spectroscopic target. This was
because, in order to acquire simultaneous observations of order 10
galaxies per exposure, we used either a movable multislit assembly
(labeled ``slitlets'' in Table 1; Geary et al.~1986) or a set of
custom-made aperture plates (Fabricant et al.~1991). The choice of
spectroscopic targets for any individual field was constrained by
factors associated with laying out the aperture plates or the
positions of the slitlets. Nevertheless the large number of aperture
plate or slitlet configurations we used (11) resulted in a reasonably
uniform sampling of cluster member galaxies.

The total exposure time was 1--2 hrs for each field. These
exposure times were built up of individual exposures of 20--30 minutes
each in order to avoid excessive contamination by cosmic rays. We
traced and extracted sky-subtracted spectra from each individual
flat-fielded frame using standard IRAF tasks. The tracing of the
spectra had typical rms residuals of $\sim$1 pixel. Then the He-Ne-Ar
arc lamp spectra (extracted from the 2-D spectral data using the same
tracing as for the corresponding object spectrum) were wavelength
calibrated by identifying at least 30 spectral lines and fitting their
pixel positions against their known laboratory wavelengths with a
third order polynomial function. The fits produced solutions with rms
residuals of $\sim$1\AA\ ($\sim$50 km s$^{-1}$). Subsequently, we
co-added the spectra from the individual frames and checked the
accuracy of the wavelength solution by measuring the positions of
several night sky lines in galaxy spectra that were not sky
subtracted. The typical differences between the measured and actual
sky line positions were generally $<$ 1\AA\ except for the June 1989
and January 1990 observations, which showed larger offsets
($\sim$20\AA). Although we were unable to determine the cause of these
large offsets, we were able to correct for both the large and small
wavelength offsets by shifting the overall wavelength scale of the
data by the average difference of the measured and true wavelengths of
the night-sky lines. We verified that the large corrections applied to
the January 1990 data were accurate by comparing the velocity of
galaxy \#7 (see Table 2) from that observing run with a previous one
and noting a small velocity difference ($\sim$68 km
s$^{-1}$). Moreover, observations of NGC 4486B during the June 1989
and January 1990 runs for calibration purposes resulted in corrected
velocities that were in general agreement (within 180 km s$^{-1}$)
with published values (de Vaucoulers et al.~1991).

The galaxy recessional velocities were computed using the IRAF task
FXCOR. This task allowed us to cross-correlate the observed spectra
(after removal of cosmic rays and any possible emission lines and
ignoring spectral regions containing bright night-sky lines) with a
high signal-to-noise spectrum of NGC4486B generated for the CfA
redshift survey. Table 2 lists the coordinates (in epoch J2000),
heliocentric velocities and errors, the Tonry \& Davies (1979) `R'
parameter (TDR) from the cross-correlation, the R-band magnitude, and
comments, as appropriate, for the 147 galaxies in our original
sample. In the top part of the table we list the 55 spectroscopic
targets that we confidently identify as galaxies, based on a high TDR
value ($\ge$3.5) and visual inspection of the spectra.  The galaxies
are ordered by right ascension and numbered sequentially.  Each
cross-correlation was further inspected and the derived redshifts were
confirmed by identifying the locations of prominent absorption line
features.  The bottom part of the table lists the remaining objects in
the original sample, also in right ascension order.  Secure
identification was not possible for these objects, either because the
object was not observed spectroscopically or because its spectrum was
insufficient to admit a positive identification (e.g., TDR values $<$
3.5).  The objects in the latter case were generally faint: the median
magnitude of the unidentified spectroscopic targets was 19.5, versus a
median magnitude of 18.6 for the identified ones.  In the bottom part
of the table we also indicate possible stars, which we consider to be
any unresolved object brighter than the BCM and in fact, two of these appear in
the HST Guide Star Catalog.  A possible background galaxy is
indicated as well.

The velocity errors listed in this table were computed by adding in
quadrature the formal uncertainty from the cross-correlation (as
output by task FXCOR), with our estimates for the errors from
sky-subtraction, arc-lamp wavelength calibration ($\sim$50 km
s$^{-1}$), and shift in the absolute wavelength scale from the
wavelengths of the night sky lines ($\sim$20 km s$^{-1}$). The errors
caused by sky-subtraction were estimated by varying the parameters of
the function used to fit the positional variation of the night sky
brightness in the cross-dispersion direction.  We estimate these
errors to be $\sim$45 km s$^{-1}$.  Finally, we note that the template
galaxy used for the cross-correlation was not completely successful at
identifying galaxies with strong Balmer absorption line features. Two
of our galaxies (numbered 9 and 11, denoted BLG in Table 2) appeared
to show such spectra. We calculated their redshifts manually and
estimate a velocity error of 150 km s$^{-1}$ for them.

As an internal check on our velocity measurements, we made repeat
observations of five galaxies during the course of the
project. The average difference in the derived velocities is 58
km s$^{-1}$, well within our quoted errors.  And as a final check, we
compared our velocities with those published by OFHF for the seven
galaxies in common, viz., 1, 2, 12, 34, 42, 46, and 49 (which
correspond to galaxies 231, 235, 225, 201, 224, 218, and 234,
respectively, in OFHF's numbering scheme). The average velocity
difference is $-51.0$ km s$^{-1}$ and the rms dispersion is 130 km
s$^{-1}$, which is also within our velocity uncertainty. The largest
difference was for galaxy 1, which showed a difference of $-$324 km
s$^{-1}$; however, OFHF report the presence of a cosmic ray in their
spectrum of this galaxy, which might have affected their estimate of
its redshift.

In order to determine how representative of the cluster our final
spectroscopic sample is, we compared the radial distributions of the
identified galaxies in the top part of Table 2 with all the galaxy
candidates in our catalog minus an estimate for the galaxy field
population (Trentham 1998). The ratio of the number of identified
galaxies to the number of candidate galaxies was computed for three
radial bins each 1\farcm2 (200 kpc) wide centered on the BCM. We
obtained values of 0.49, 0.5, and 0.46 for the ratios in the three
bins. This distribution is fairly flat, which suggests that there is
not a serious radial bias in the final sample of identified cluster
galaxies.  We also compared the number-magnitude counts of galaxies in
the velocity range 50,000--60,000 km s$^{-1}$, which are likely
cluster members, with the photometric results of Trentham (1998).  Our
sample of members of A665 is not complete; we have spectroscopically
confirmed the membership of only 57\% of the expected number of 
cluster galaxies down to $m_R = 19$ (or 40\% of the expected members 
brighter than $m_R = 20$). Nevertheless the shape of the number-magnitude
distribution is quite similar to that of Trentham's at least down to
magnitude 19.  Thus although our sample of galaxies is not a complete
or well-defined subsample of the full population of cluster members,
it appears fairly representative.

%It is not intended that the preceding discussion be
%taken as definitive, but rather our goal is to show that the
%identified galaxy sample is representative, both in magnitude and
%position on the sky, of the overall galaxy population in A665.

\subsection{X-ray}

Our \rosat\ high resolution imager (HRI) observation of A665 was
carried out in two parts; the first set of observations (rh800774) was
done on April 1996 and lasted for a total of 41,224 seconds (live-time
corrected) while the second set (rh800900) was performed in May 1997
and contained 57,650 seconds of live-time corrected exposure. These
individual observations were summed after verifying that there was no
significant offset in their relative pointing directions. Next, we
boresight-corrected the absolute positions of the combined data set
using two known radio sources (MB20 and MB28 from Moffet \& Birkinshaw
1989) that were also X-ray sources. The difference in X-ray and radio
source positions were 0\farcs4 (right ascension) and 2\farcs7
(declination). With these corrections, we estimate that the absolute
X-ray positions are accurate to $\lsim$1$\arcsec$.

The raw data were block-averaged to produce an image with
8$\arcsec$ pixels. The total background level was determined by
computing the counts from an annular region centered on the cluster
with an inner radius of 5\arcmin\ and a width of 10\arcmin\ avoiding
obvious point sources.  This yielded a background level of $4.8 \times
10^{-3}$ counts s$^{-1}$ arcmin$^{-2}$ which is consistent with the
range seen in other ROSAT HRI observations (David et al.~1998). Note that
the total number of background subtracted counts within a radius of 5$\arcmin$ 
of the peak of X-ray emission ($\alpha_{\rm
J2000} = 8^{\rm h} 30^{\rm m}59.8^{\rm s}$, $\delta_{\rm J2000} =
65^\circ50^\prime31.3^{\prime\prime}$) is $\sim$ 17780. Figure
1 shows an overlay on the R-band image of the X-ray surface brightness
contours from the background-subtracted HRI data adaptively-smoothed
to an approximate signal-to-noise ratio of 5 which yields a typical smoothing scale of 
15$\arcsec$ within the inner 2.$\farcm$5 of the image (see Huang \& Sarazin
1996 for a description of the smoothing algorithm).

\section {DATA ANALYSIS}

\subsection{Optical Data Analysis}

We assembled a sample of 89 potential cluster members by combining our
55 galaxies with 34 different ones from OHFH. Note that
OHFH's entire sample contained 41 galaxies, but there were seven objects 
in common with our sample; for these we use our redshift measurements.
Cluster membership was determined using an iterative 3$\sigma$
clipping criteria (Yahil \& Vidal 1977). The outlier galaxies were
easily identified (after a single iteration): we find the same eight
outlier galaxies as OHFH did in their sample as well as four more in
our own sample. These latter are flagged in Table 1 as lying in either
the foreground or background. We are left with 77 galaxies as cluster
members in the total sample. Due to the relatively modest number of
galaxies in our sample, we replaced the classical statistical
estimators for the mean and dispersion with the more robust biweight
estimators (Danese et al.~1980, Beers et al.~1991). The redshift of
the cluster from the total sample is $0.18285^{+0.00045}_{-0.00064}$
and the 1-D velocity dispersion is $1390^{+120}_{-110}$ km s$^{-1}$ (1
$\sigma$ errors). The uncertainties were determined through a
bootstrap technique using the ROSTAT software (Beers et al.~1991).
Our value of the cluster's velocity dispersion is consistent 
with the value derived from the well-known correlation between X-ray
temperature $kT$ and velocity dispersion $\sigma$.  Using an average
X-ray temperature for A665 of $kT = 8.3$ keV (Hughes \& Tanaka 1992)
and the relationship between $kT$ and $\sigma$ given by Girardi et
al.~(1996), we estimate $\sigma$ = 1230 km s$^{-1}$.
%
% sigma = 10^{2.53 \pm 0.04} T^{0.61\pm0.05}
%
Table 3 lists the redshift and 1-D velocity dispersion for the total
sample and various subsamples of the cluster galaxies.  Since we are
especially interested in the dynamics of the galaxies located near the
center of the cluster, we concentrate the remainder of our analysis on
the 54 galaxies located within 4\farcm5 (750 kpc) of the BCM (galaxy
\#34 in Table 2). Figure 2 shows the velocity histogram for these
galaxies.

We applied a number of statistical tests to search for evidence of
substructure in the spatial and kinematical distribution of the
central galaxies in A665. Our tests find weak evidence for skewness in
the velocity distribution.  Specifically, using the B1 (D'Agostino 1986) 
and W (Yahil \& Vidal 1977) tests we can reject the Gaussian hypothesis 
at the 92\% confidence level. The B1 test is the usual third moment in the 
distribution while the W test computes the skewness based on ordered data.  
Note that the more conservative Asymmetry Index estimator (Bird \& Beers
1993), which tests for symmetry by comparing gaps on the left and 
right side of the ordered velocity data, yields a significance of 
$\sim$97\% for skewness. The statistical significance of other deviations from
Gaussianity (e.g., kurtosis and bimodality) is even smaller.  Therefore, the
velocity distribution is barely distinguishable from Gaussian in form.

Furthermore there is no evidence for significant substructure in the
2-D spatial distribution of the centrally located galaxies, i.e., we
do not detect strong asymmetries or bimodality in the locations of the
galaxies. 

Finally, we tested for the presence of substructure using 3-D tests
that consider both the spatial and kinematical positions of the
galaxies. The only test that resulted in even a marginal signal is the
Dressler-Schectman test (Dressler \& Shectman 1988). This test
compares the local mean velocity ($v_i$) and dispersion ($\sigma_i$)
for galaxy $i$ (computed in this case from the seven nearest
neighbors) with the global mean velocity ($v$) and dispersion
($\sigma$) by calculating the quantity $\delta_i$, where $\delta_i^2 =
[(\sqrt N +1)/\sigma^2][{(v_i - v)}^2 + {(\sigma_i - \sigma)}^2]$ and
$N = 55$ is the number of sample galaxies, and then summing to compute
the statistic, $\Delta = \sum \delta_i$.  The statistical significance
of the value $\Delta= 69.2$ that we calculate for the galaxies in A665
was determined through Monte-Carlo simulations. We find that our
observed value of $\Delta$ is larger than 95\% of the $\Delta$ values
obtained by random simulations in which the velocities were shuffled
while keeping the positions fixed.  Figure 3 is a graphical
representation of the Dressler-Schectman test, where a circle, whose
radius is proportional to the quantity $e^{\delta_i}$, is plotted at
the position of each cluster member galaxy. A concentration of large
circles in any particular area indicates the presence of galaxies
whose kinematical properties differ from the global values.  Although
no obvious pattern dominates the deviations between the local and
global values, the plot shows that most of the galaxies with large
delta values (i.e., large circles in Fig.~3) lie close to a line that
is perpendicular to the axis of X-ray elongation.  These galaxies have
local mean velocities that differ from the global mean velocity by as
much as 1350 km s$^{-1}$ and signal the presence of non-relaxed
groupings of galaxies in the cluster. In the following section we will
review what implications can be drawn from these results.

We do not detect any significant peculiar velocity of the BCM with
respect to the sample of 54 centrally-located galaxies. In particular
we measure a peculiar velocity of $20 \pm 220$ km s$^{-1}$ while OHFH
found a velocity of 447 km s$^{-1}$ for the BCM relative to their
sample of galaxies, a difference that was statistically significant at
the 98\% confidence level. Our separate, independent velocity
measurements for the BCM are in agreement, so it is the difference in
our estimates for the mean recessional speed of the cluster that must
be the cause of the discrepancy.  As Table 3 shows, the galaxies
located close to the center of the cluster (our sample) have a mean
velocity that is slightly larger than the mean velocity of the
galaxies located in the outer regions of the cluster. Since the mean
recessional velocity of the latter sample agrees well with OHFH's
sample, the discrepancy between our results, at least, is
explained. The relativistically-corrected velocity difference between
the inner and outer samples of galaxies is $580\pm290$ km s$^{-1}$,
yet another marginal (2 $\sigma$) indication that the galaxy
population in A665 is not fully relaxed.  Whether this indicates that
there is a radial velocity gradient in the cluster, or a velocity
differentiation based on galaxy type awaits improved imaging data and
a larger sample of measured redshifts.

Finally, Figure 4 shows a grey scale map of the surface density
of central cluster galaxies overlaid with the X-ray surface
brightness contours. Two features to note are (1) the offset of the
peaks of the X-ray emission and the galaxy density by a small amount
$\sim$1\arcmin\ (167 kpc) and (2) the NW-SE elongation of the brighter
parts of both distributions. The generally good agreement between the
spatial distributions of the gas and confirmed cluster member galaxies
argues strongly against the possibility that the distorted appearance
of the X-ray emission is the result of a superposition of unrelated
clusters along the line of sight.

\subsection{X-ray Data Analysis}

We carried out an elliptical isophotal analysis of A665's X-ray
surface brightness using the algorithm developed by Jedrzejewski
(1987) as implemented in IRAF (using task ellipse in the stsdas
package). The program uses an iterative least-squares technique to
model the radially decreasing surface brightness with elliptical
isophotes. The centroid (row and column pixel positions), axial ratio,
and position angle of the major axis are varied to obtain the best
fit.  Figure 5 shows the results of these fits to the
adaptively-smoothed \rosat\ HRI data in four panels that present the
radial variation from the peak of the X-ray emission ($\alpha_{\rm
J2000} = 8^{\rm h} 30^{\rm m}59.8^{\rm s}$, $\delta_{\rm J2000} =
65^\circ50^\prime31.3^{\prime\prime}$) of the axial
ratio, position angle (measured positive counter-clockwise from
north), and centroid shifts in right ascension and declination.

The errors shown were estimated using a Monte Carlo technique. One
hundred simulated images of the cluster were generated by adding
Poisson noise to the original adaptively-smoothed X-ray data. Each
separate realization was re-smoothed and then run through the ellipse
program.  The error at each radial point was estimated by computing
the rms dispersion of the 100 fitted values obtained from the
different realizations. Note that this technique explicitly takes
account of the statistical error from Poisson noise as propagated
through the ellipse fitting process.

As figure 5 shows, there is significant centroid movement, ellipticity 
variation, and isophotal twisting. For example, by a radius of 2\farcm5
the centroid of the isophotes has shifted by over $1^\prime$ toward
the NW. Note that this shift is larger than the overall smoothing 
scale within this region ($\sim$ 15$\arcsec$) and roughly the same 
as the direction of elongation of the
central part of the galaxy surface density distribution mentioned in
the previous section.  The shape and orientation of the isophotes are
also a function of radius. They are initially fairly elliptical
($\epsilon = 1-b/a \sim 0.17$) with the major axis aligned slightly
toward the NE. Over the next arcminute in radius, the isophotes twist
as the position of the major axis rotates by $-45^\circ$ (i.e., toward
the west).  At this point the major axis points in nearly the same
direction as the direction of the centroid shift.  The orientation of
the isophotes remains nearly constant for the next arcminute or so in
radius before continuing to twist further toward the west.  Beyond a
radius of $\sim$2\farcm5, the signal to noise ratio of the data has
become too low to pursue further fits. These results are generally
consistent with the substructure detected in the PSPC images of A665
by Hughes \& Birkinshaw (2000), which required a number of spatial
components, at least two isothermal-$\beta$ models as well as two
ellipsoidal distributions of hot gas, to explain in detail. 
 
\section {DISCUSSION}

Based on the results of our optical data alone, we would be hard
pressed not to conclude that A665 is a relaxed cluster.  The velocity
dispersion, although high, is not inconsistent with measurements of
other rich clusters (e.g., Zabludoff et al.~1993; Edge \& Stewart
1991). We detect only marginal evidence for substructure and
non-Gaussianity in the velocity distribution of the cluster
members. The BCM appears to be at rest with respect to the
gravitational potential, at least as traced by the dense concentration
of galaxies near the cluster center.  In short, all the tests that we
have used on the galaxy data are consistent with a fairly simple,
relaxed cluster scenario.

On the other hand, the morphology of the X-ray emitting gas indicates
an entirely different situation for the dynamical state of the
cluster.  The strongly asymmetric distribution of the X-ray emitting
gas, the isophotal twisting and centroid variation presented above,
and the exceptional nature of A665's morphology as discussed by Buote
and Tsai (1996) all argue strongly against a typical relaxed cluster
interpretation.  These facts suggest that the hot cluster gas is not
in hydrostatic equilibrium with the cluster gravitational potential
or, if it is, then the gravitational potential must be strongly
asymmetric, clumpy, or highly structured in some way.  The
possibility of a chance line-of-sight superposition of two isolated,
nearly-regular, clusters giving rise to the distorted appearance
of A665 can be rejected based on the general agreement between the
spatial distribution of cluster member galaxies (with $cz \sim 50000
\ \rm km \, s^{-1}$) and the X-ray emission (Fig.~4). An alternate
scenario 
that may be consistent with the observations is one in which A665 is
undergoing a recent merger and that the inherent complexities in the
dynamics of the galaxies are somehow hidden from us.  This scenario
could naturally account for the X-ray substructure and, depending on
the merger epoch and viewing geometry, might be consistent with the
spatial and kinematical properties of the galaxies. In the following,
we explore this merger hypothesis in more detail.

Numerical simulations (e.g., Evrard 1990; Pearce et al.~1994;
Schindler \& M\"uller, 1993; Roettiger et al.~1996) show that during a
major merger, the gravitational potential of a cluster is severely
disrupted. Shocks heat and compress the gas, eventually redistributing
a portion of the energy associated with the merger to the cluster gas.
However during the actual merger itself, the gravitational potential
evolves too rapidly for the cluster atmosphere to respond on the
sound-crossing timescale, and consequently the gas falls out of
hydrostatic equilibrium. Shocks, the exchange of gas between the
merging components, and the reaction of the gas to the gravitational
forces all give rise to structures, elongations, and other asymmetries
in the X-ray emission.  These processes, we believe, provide a
plausible qualitative explanation for the X-ray substructure seen in
A665.  Yet these simulations also predict the presence of substructure
in the spatial and kinematical distributions of the cluster member
galaxies.  Have we failed to observe these effects in A665 because 
our galaxy sample is too small, the properties of the merger (epoch
and viewing geometry) are unfavorable (see, for example, Pinkney et
al.~1996), or because the underlying premise of a major merger in A665 is
just plain wrong?

We decided to test the merger hypothesis by comparing our optical data
with the results of N-body simulations of simple head-on cluster
mergers. We do not aim for a precise match between the data and
simulations (and we do not carry out an exhaustive study); we try only
to determine whether this general type of merger can produce the subtle 
kinematic deviations from an apparently relaxed cluster that we
observe. We assume that the galaxies are good probes of the underlying
gravitational potential and that light traces mass.

Given the significant distortions in the X-ray emission, we have
chosen to focus on major mergers for maximum effect. We tried two
general types of merger: one consisting of a merger of equal mass
components (1:1) and one with merging components in mass ratio of 1 to
2 (1:2). All simulations were performed using Hernquist's N-body code
(TREECODE, Hernquist 1987). We started with simple, idealized initial
conditions.  Each of the component subclusters was modeled as an
isothermal sphere (King 1966, Binney \& Tremaine 1987) characterized
by a concentration parameter of 1.08 (so that the tidal radius is
$\sim$ 12 times the core radius). The initial core radii were fixed at
a value of 250 kpc for the equal mass cluster simulation, while in the
1:2 merger they were 250 kpc (main cluster) and 198 kpc (subcluster).
These choices ensured an equal matter density between the merging
subclusters and are consistent with values found in the literature
(e.g., Mohr et al.~1999). Note that this choice fixed the length scale
for the simulations and left, as the single remaining scalable
parameter, either the mass per N-body particle or the time step of the
simulation.  The two subclusters were allowed to merge head-on under
the influence of their mutual gravity.  At the start the two merging
components were separated by 8 Mpc and 6 Mpc for the 1:1 and 1:2
models, respectively, and had initial velocities consistent with their
free-fall velocities ($\sim 1000 \ \rm km \, s^{-1}$).  The total
number of particles in the computations was 30000, divided
appropriately between the two components.

In order to compare to the observed velocity distribution of the
galaxies it was necessary to apply a numerical scaling to the
velocities of the N-body particles, since the model calculations were
done in scale-free coordinates. Fixing the velocity scaling is
equivalent to defining the time step of the simulations which in turn
determines the masses of the merging components. The other quantities
of interest to us for the comparison were the merger epoch and the
viewing direction, i.e., the angle between our line-of-sight and the
merger axis.  We investigated three different viewing angles:
30$\arcdeg$, 45$\arcdeg$, and 90$\arcdeg$ (where 0$\arcdeg$
corresponds to viewing along the merger axis).  Only N-body particles
within a projected radius of 750 kpc about the center of mass of the
system were used in the model velocity distribution. The
Kolmorov-Smirnoff (KS) test was used for the comparison.

The results are shown in Figure 6. The two panels show the two
different mass ratios used for the initial conditions.  In each panel
the vertical axis shows the initial mass of the main merging
subcluster while the horizontal axis shows the time since
core-crossing. The different symbols indicate the different viewing
directions assumed. Only models with a (2-sided) KS probability for
rejection of 90\% or less (i.e., those with a maximum difference
between the model and the data distributions of $\lsim$0.076) are
plotted.  Overall 8700 possible models as a function of epoch (from
$-$6 Gyr to 5 Gyr with a typical timestep of $\sim$0.3 Gyr) and
initial main subcluster mass (from $0.7\times 10^{15}\, M_\odot$ to
$1.5\times 10^{15}\, M_\odot$ with a typical mass spacing of
$\sim$$4\times 10^{13}\, M_\odot$ ) were sampled for each viewing
angle. These results demonstrate that the observed velocity
distribution in A665 is consistent with the velocity distribution
expected from the major merger of two similar sized subclusters, close
to or after the time of core-crossing, with the exact epoch depending
on the mass ratio and viewing geometry.

Another view of the mass of A665 inferred from the allowed merger
models is shown in Figure 7.  Here we plot the mass within $r_{500}$,
which is the radius at which the cluster density is equal to 500 times
the critical density. Evrard et al.~(1996) have shown from numerical
simulations of cluster evolution that $r_{500}$ can be expressed as a
power law of the gas temperature. For A665 with a global temperature
of 8.3 keV, we compute a value of $r_{500} = 1.5$ Mpc from this
relation.  According to Fig.~7 the inferred total cluster mass within
this radius based on galaxy dynamics is in the range $1.6\times
10^{15}\, M_\odot$ to $2.4\times 10^{15}\, M_\odot$.  These estimates
are consistent with the total mass ($1.7\times 10^{15}\, M_\odot$)
predicted by the scaling relations proposed by Evrard et
al.~(1996). However, they are somewhat in excess of the mass estimate
($0.9-1.5\times 10^{15}\, M_\odot$) from an X-ray analysis by Hughes
\& Tanaka (1992) under the typical assumptions of hydrostatic
equilibrium and spherical symmetry.  Numerical simulations of cluster
mergers and evolution (e.g., Evrard et al.~1996, Roettiger et
al.~1996) show that one can overestimate or underestimate the actual
mass of a merging cluster by factors of 2 if one assumes hydrostatic
equilibrium and isothermal $\beta$-models. The exact difference
between the true cluster mass and the inferred mass from X-ray
analysis depends on the geometry, epoch, and other gas parameters
(e.g., temperature, core radius, $\beta$) of the merging
clusters. Thus, it is not surprising that our mass estimates from
galaxy dynamics are not in perfect agreement with the X-ray--derived
mass.

Looking at the results of the 1:1 merger in more detail, we find that
the K-S test rejects mergers at epochs slightly earlier than $-$0.5
Gyr because the velocity distributions tend to be either bimodal or
very broad (depending on the viewing angle) and therefore incompatible
with the data. On the other hand, mergers at even earlier epochs
(before $-1$ Gyr) or at late epochs (after 1 Gyr) are rejected because
the K-S test finds the velocity distributions to be more symmetric
and/or wider than our data. Thus, this test is more sensitive to the
marginal non-Gaussianity of our data than the generic statistical
tests that we used before (see section 3.1).  The situation for the
1:2 mass ratio mergers is somewhat different.  The velocity
distributions produced by these simulations do allow a larger range of
models due to the smaller effect that the subcluster has on the
overall shape of the velocity distribution. In other words, our
analysis is consistent with two similar-size clusters caught in the
middle of a merger or a larger system being affected by a smaller
subcluster. Note that there are no acceptable models for earlier
epochs (more than 1 Gyr before core-crossing) that would correspond to
relaxed clusters seen in projection if viewed along the merger axis
since such models would display spatial distributions of galaxies that
would be clearly bimodal and therefore inconsistent with the data.

Our optical velocity data showed marginal evidence for a difference in
the mean velocity and velocity dispersion with position (Table 3). We
have used this information in an attempt to further discriminate among
allowed models. From the N-body models we extracted the line-of-sight
velocities of 54 objects in the inner 750 kpc (corresponding to our
central sample) and 23 objects from within an annular region from 750
kpc to 2 Mpc (corresponding to the outer sample).  Here we have
estimated 2 Mpc to be roughly the outermost radial extent of OHFH's
study.  For each merger epoch, mass ratio, and viewing angle we
calculated the mean velocities and dispersions in the two extraction
regions.  Sampling errors were taken into account by averaging the
values obtained from 100 independent random extractions.  The 1:1
merger does not show any evidence at all for a gradient in the mean
velocity, although there is evidence for a velocity dispersion
gradient especially for epochs close to core-crossing.  Specifically,
the velocity dispersions from the two regions differ by approximately
300--600 km s$^{-1}$ with the maximum difference occurring at core
crossing.  The situation is much different for the 1:2 merger.  In
this case there is evidence for a gradient in both mean velocity and
velocity dispersion. For epochs between $-$1 Gyr and 2 Gyr from core
crossing, the mean velocities differ by 200--400 km s$^{-1}$ and the
velocity dispersions differ by 400--800 km s$^{-1}$ for the inner and
outer galaxy samples.  These values are comparable to our measured
values. Therefore, of all the models considered in our analysis, the
1:2 merger seen close to the time of core crossing appears to be the
most consistent with our galaxy velocity data.

Although detailed modeling of the properties of the X-ray emission
arising from a merger of this type is beyond the scope of our work,
the properties of the merger inferred from the galaxy velocities
(i.e., small mass ratio, close to core-crossing) are qualitatively
consistent with the distorted morphology of the X-ray emission
(Roettiger et al.~1996). We await upcoming X-ray spectral images from
\chandra\ that will allow us to further investigate the properties of
the hot gas in A665 and probe the evolutionary state of this complex
merging system in greater detail.

\section {SUMMARY AND CONCLUSIONS}

We measured new R-band optical magnitudes for 147 galaxy candidates in
the vicinity of the rich galaxy cluster A665.  A total of 89 of these
candidates were observed at the MMT; good signal-to-noise spectra that
resulted in the identification and determination of recessional
velocities were obtained for 55 galaxies. Combining with data in the
literature results in a total of 77 known cluster member galaxies in
A665.  We concentrate our study of the cluster's kinematics on the
subsample of 54 galaxies within the inner 4\farcm5 (750 kpc) central
region of the cluster.

We find, at most, marginal evidence for kinematic structure and
non-Gaussianity in the velocity data of the central subsample of
galaxies. In addition for these galaxies there is only weak evidence
for 3-D spatial and kinematical clustering as detected by the
Dressler-Schectman test. Comparison of the central subsample to the
sample of cluster members beyond 4\farcm5 of the center, shows
marginal ($\sim$2 $\sigma$) evidence for a drop in both the mean
velocity and velocity dispersion.  Taken at face value the optical
velocity data therefore appear consistent with a massive relaxed
cluster, exhibiting only subtle signs of substructure.  Others have
pointed out that the spatial distribution of galaxies shows some
evidence for substructure.

Our deep \rosat\ HRI observation of A665 reveals strong signatures of
substructure in the spatial distribution of the X-ray emitting gas
that are consistent with PSPC observations (Hughes \& Birkinshaw
2000). We have measured centroid shifts, ellipticity variations,
and the rotation (or twist) of elliptical isophotes as a function 
of distance from the cluster center. These indicate recent merger activity
in the cluster.

In order to reconcile these two apparently conflicting views of the
cluster's evolutionary state, we have undertaken simple N-body
simulations of head-on cluster mergers. We find that the velocity
distributions produced by the merger of two subclusters with mass
ratios of 1:1 or 1:2 near the time of core-crossing provide an
acceptable match to the observed velocity distribution of the central
subsample for a range of reasonable viewing geometries.  In addition,
near the epoch of core-crossing the 1:2 merger produces a radial
gradient in mean velocity and dispersion that also agrees with our
measurements.  A major merger of this type is at least qualitatively
consistent with the distorted X-ray morphology of A665.

Major new insights into the nature of this cluster and the process of
cluster formation in general should be forthcoming with the data
expected from the new generation of X-ray missions recently launched. \chandra\ and \xmm\ will 
provide detailed
measurements of the cluster's gas temperature and density that are
sure to provide excellent views of shocks and other plasma processes
in the hot gas. Furthermore, our simulations indicate
that increasing to $\sim$ 250 the number of cluster member galaxies in 
A665 with good redshift measurements would allow us to
observe directly the effects of the merger on the galaxy 
velocity distribution.  Each of these techniques yields a 
different view of the ongoing merger in A665 and thus merits follow-up.

\acknowledgments

This work was partially supported by a PPARC grant to MB and NASA
grants NAG5-3432, NAG5-4794, and NAG5-6420 to JPH. We would like to
express our thanks to Neil Trentham who kindly provided us with his
A665 optical images and to Tad Pryor for insightful comments about the
work.  We also thank the CfA time allocation committee for their
patience and generosity in awarding us MMT observing time over the
many years this project took. Support from Monique Arnaud and the
hospitality of the XMM group at the Service d'Astrophysique of the
CEA-Saclay is also gratefully acknowledged by JPH.

\newpage

\clearpage

\figcaption[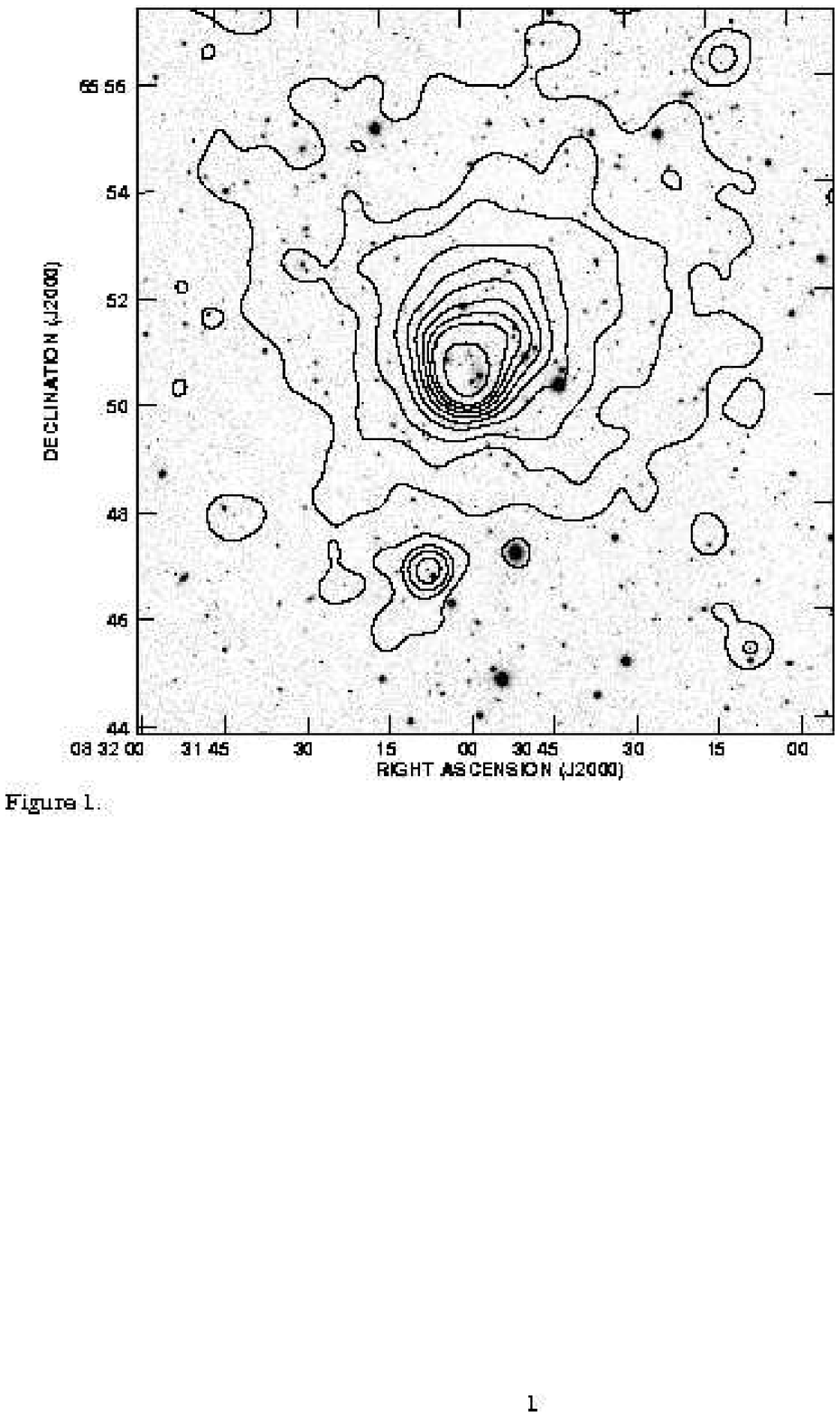]{Overlay of X-ray surface brightness contours on an
optical R band grayscale image of A665. The contour levels are at
values of 6, 18, 30, 48, 66, 72, 102, 120, 138, and $180 \times 10^{-4}$
HRI counts s$^{-1}$ arcmin$^{-2}$.  The \rosat\ HRI X-ray image has been
adaptively smoothed and clearly shows a cluster with asymmetric
isophotes.\label{Figure 1}}

\figcaption[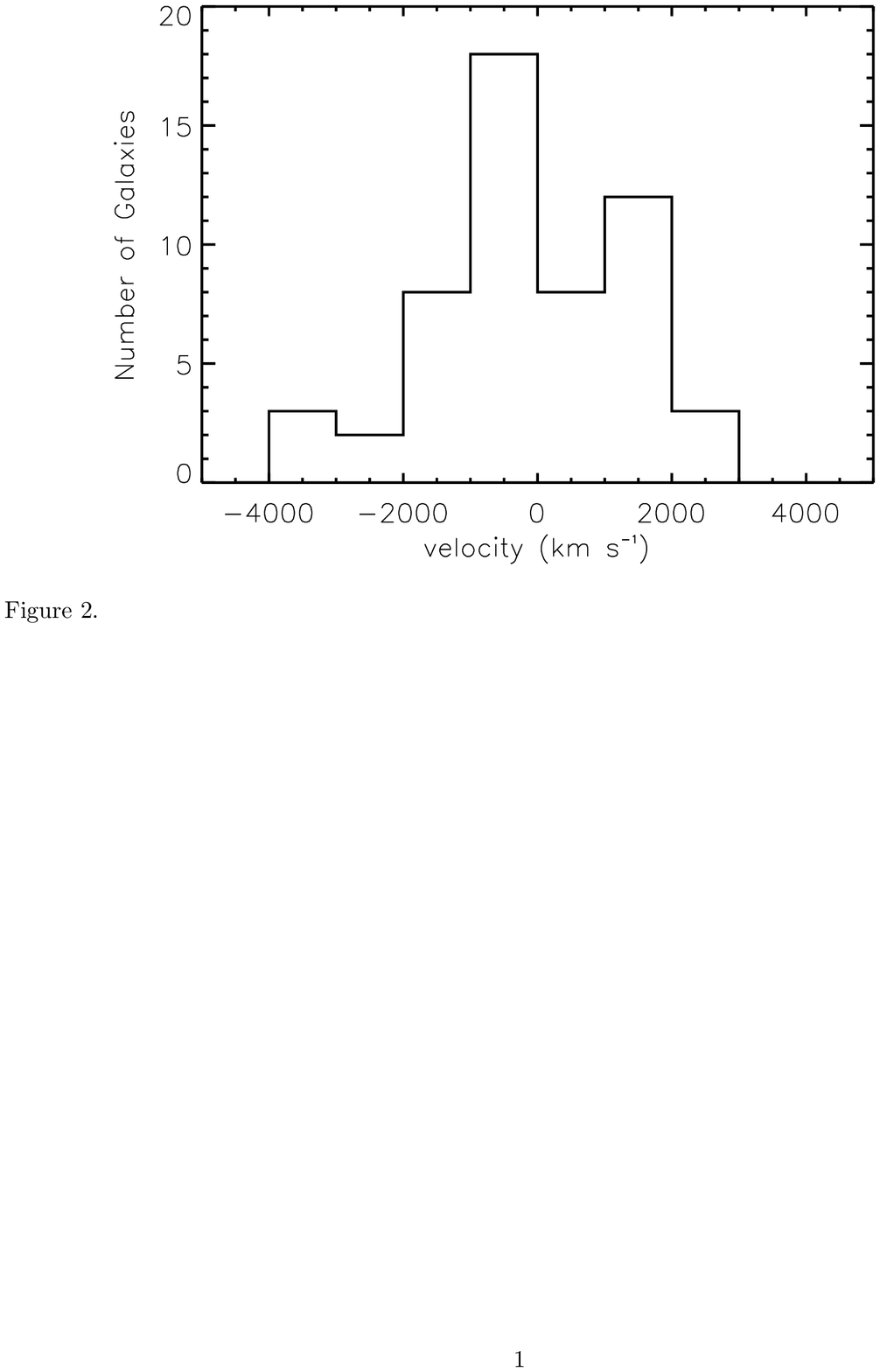]{Velocity histogram of the 54 cluster member
galaxies located within 750 kpc of the center of A665. All velocities
have been corrected to the cluster reference frame. The bin size is
1000 km s$^{-1}$.\label{Figure 2}}

\figcaption[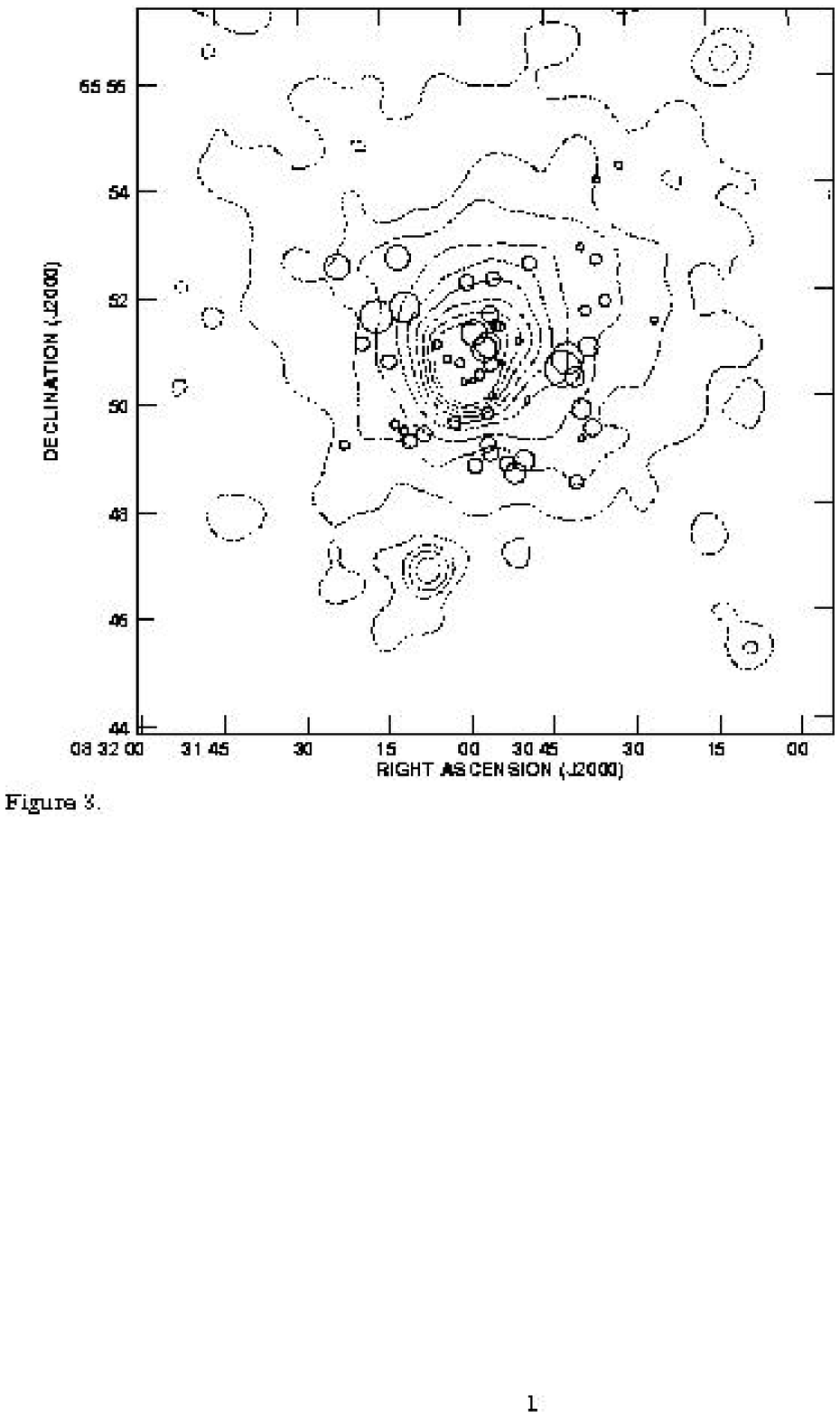]{Contours of \rosat\ HRI X-ray surface brightness
superposed with the results of the Dressler-Schectman 3-D test for
substructure. The contours are identical to Fig.~1 except that
the central contour is omitted for clarity. Each circle represents
the position of a galaxy and the size of the circle is proportional to
the exponential of the deviation between the local and global mean
velocities and velocity dispersions.\label{Figure 3}}

\figcaption[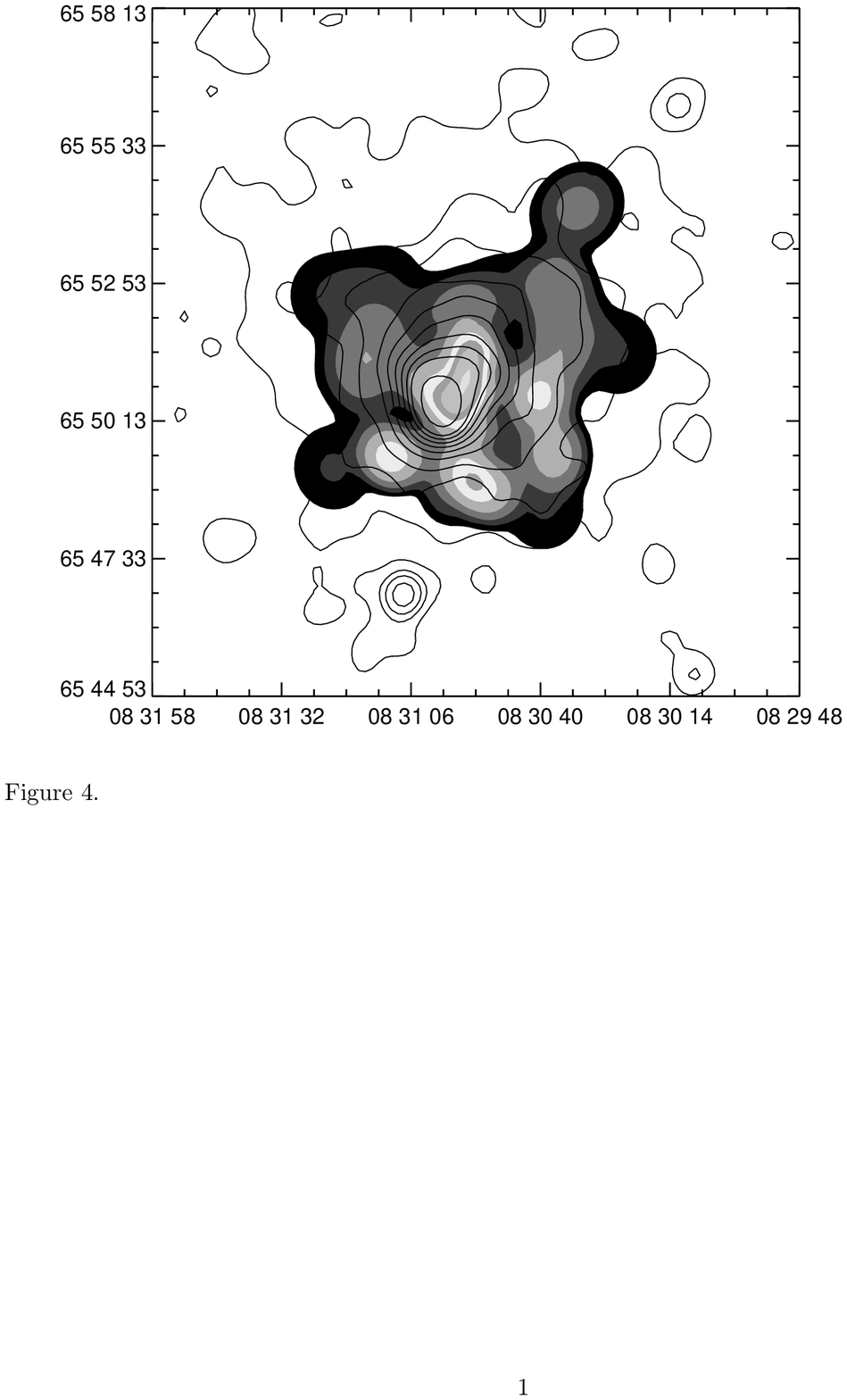]{X-ray surface brightness contours overlaid on a
grayscale map of the galaxy surface density for secure cluster members
in A665. The contours are identical to those in Fig.~1.  Both maps
have been adaptively smoothed. Note the offset in peak density between
the two distributions and the NW-SE elongation of each.\label{Figure
4}}

\figcaption[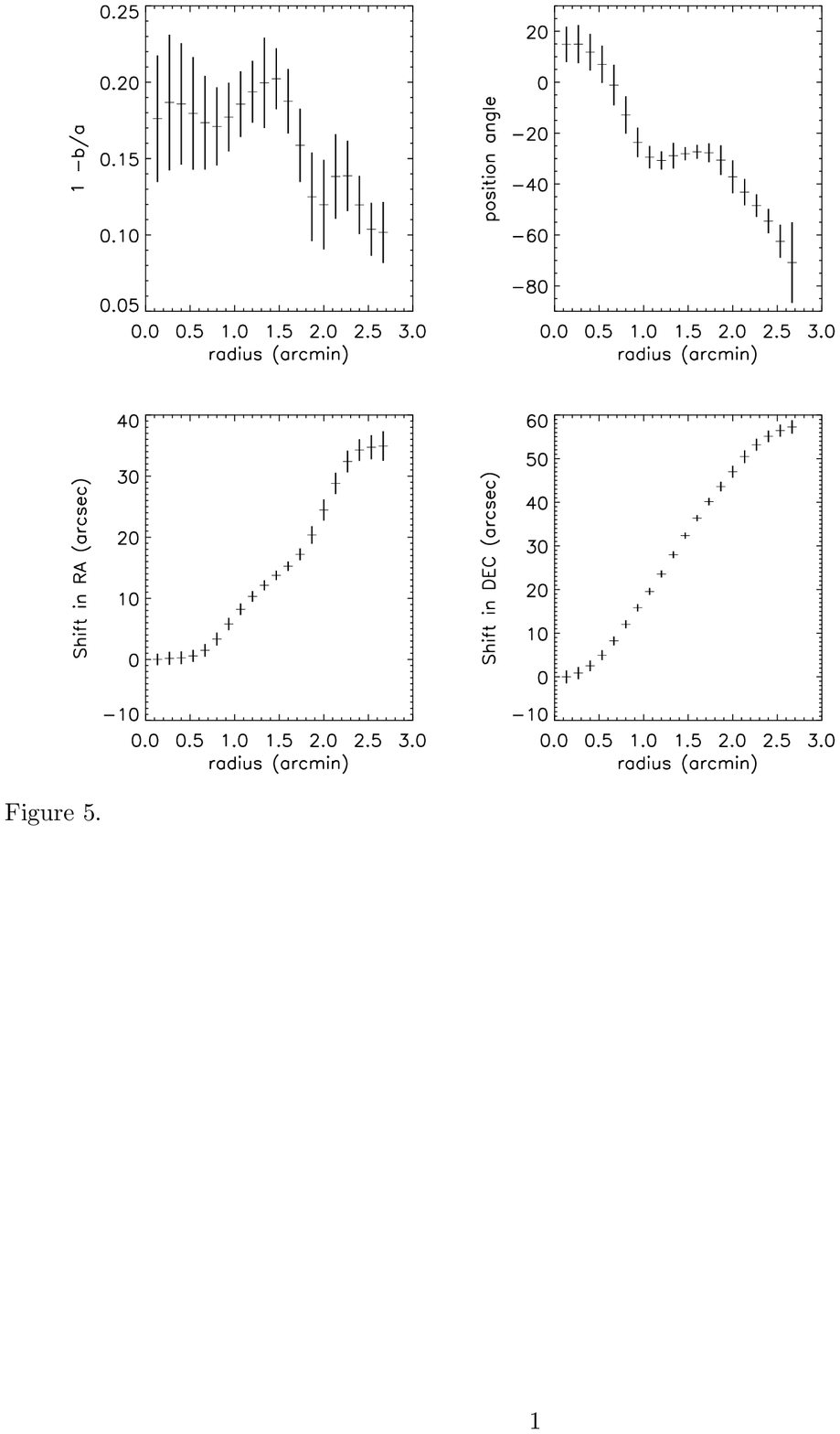]{Elliptical isophotal analysis of the \rosat\ HRI
X-ray image. The four panels (starting at top left and proceeding
counterclockwise) show the ellipticity ($b/a$ is the axis ratio),
position angle of the major axis (measured positive counterclockwise from
north), centroid shift in declination and centroid shift in right
ascension.  Note the large centroid shift apparent in the bottom
panels. \label{Figure 5}}

\figcaption[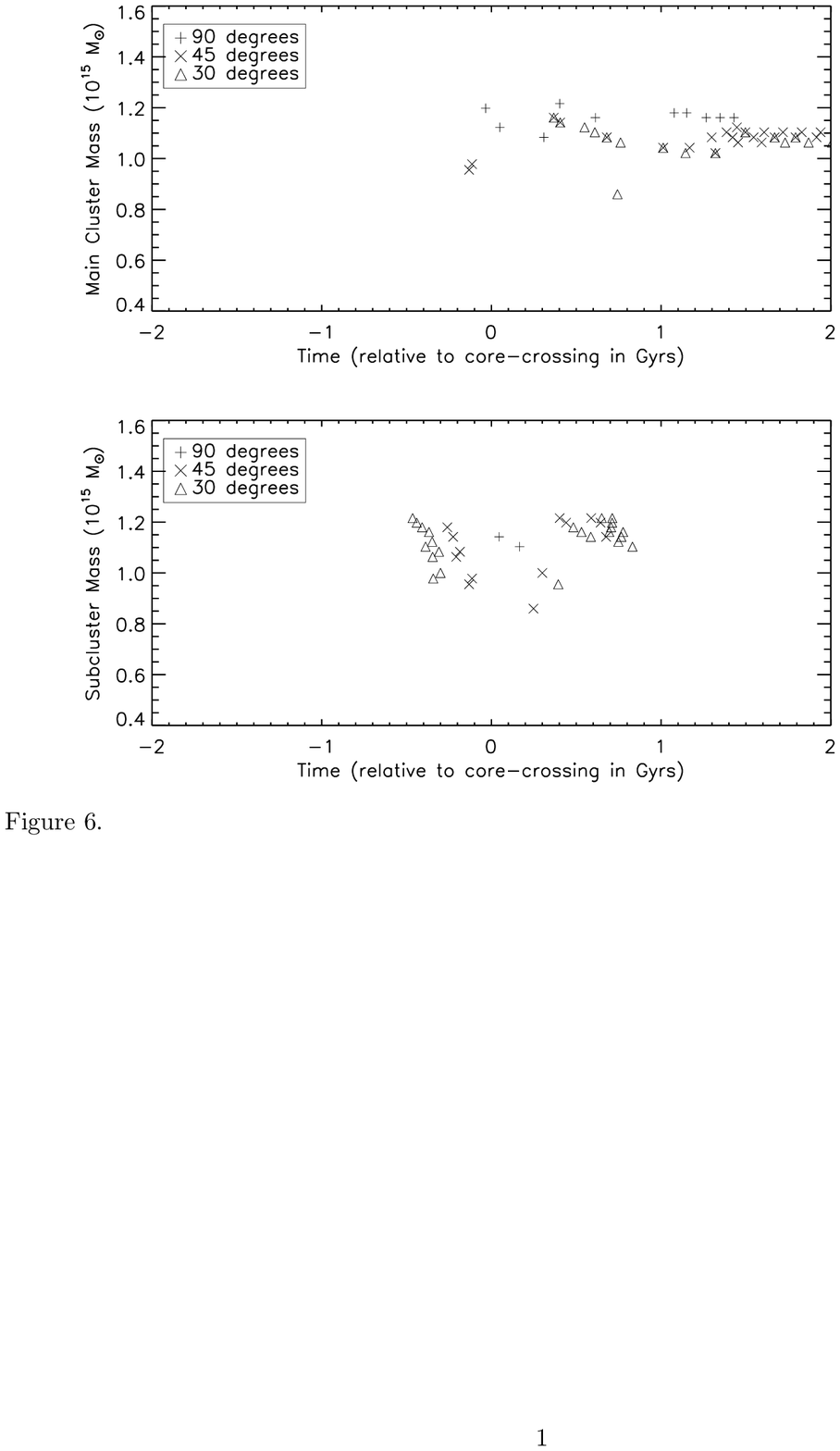]{Allowed merger models plotted as a function of main
cluster mass and time since merger, based on comparing the
line-of-sight velocity distributions from the 54 galaxies near the
center with the modeled distributions.  Those models with a K-S
probability gretaer than 90\% are shown; many other possible models
were rejected.  The top panel corresponds to a mass ratio of 1 to 2
between the merging components, while the bottom panel corresponds to
the merger of equal mass subclusters.  We have indicated how the
results depend on viewing angle for three different values (where
0$\arcdeg$ corresponds to viewing along the merger axis). In both
panels the plus symbol corresponds to a 90$\arcdeg$ viewing angle, the
cross corresponds to a 45$\arcdeg$ viewing angle, and the triangle
corresponds to a 30$\arcdeg$ viewing angle.\label{Figure 6}}

\figcaption[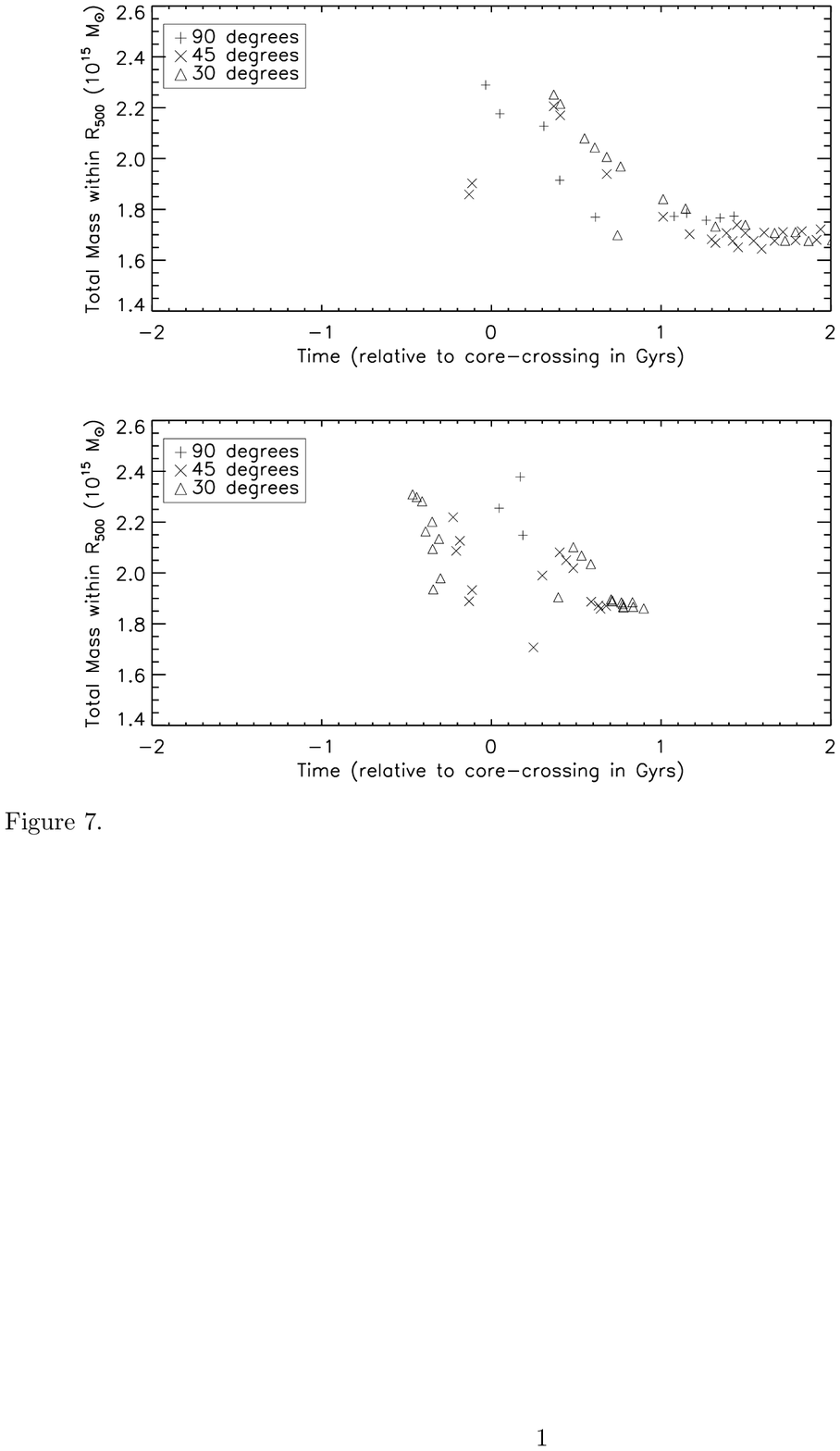]{Same as Fig.~6 except that the allowed merger
models are plotted as a function of total cluster mass within the
virial radius ($r_{500}$ = 1.5 Mpc).\label{Figure 7}}

\clearpage

\begin{deluxetable}{ccc}
\tablenum{1}
\tablewidth{0pt}
\tablecaption{Observation Log}
\tablehead{
\colhead{Dates of Observation} &
\colhead{Technique} &
\colhead{Instrument and Grating}}
\startdata
4-9 April 1986  &       slitlets &        FOGS 400 lines mm$^{-1}$\nl
20 January 1988 &       slitlets &        FOGS 400 lines mm$^{-1}$\nl
5-6 January 1989&       aperture plates & FOGS 300 lines mm$^{-1}$\nl
3-5 June 1989   &       aperture plates & Red Channel 270 lines mm$^{-1}$\nl
1  January 1990 &       aperture plates & Red Channel 270 lines mm$^{-1}$\nl
30-31 March 1990&       aperture plates & Red Channel 270 lines mm$^{-1}$ \nl
19-20 January 1991&     aperture plates & Red Channel 270 lines mm$^{-1}$\nl
\enddata
\end{deluxetable}

\begin{deluxetable}{cccccccc}
\tablenum{2}
\scriptsize
\tablewidth{0pt}
\tablecaption{Galaxies in the A665 field}
\tablehead{
\colhead{Galaxy ID}           & 
\colhead{RA(2000)}      &
\colhead{DEC(2000)}          &
 \colhead{Heliocentric Velocity}  & 
\colhead{error}          &
 \colhead{TDR value}   &
 \colhead{R magnitude}  &
\colhead{comments}}
\startdata
$\phantom{00}$1 &  8 30 34.5 & 65 51 50.0 &  54645 & $\phantom{0}$96 & $\phantom{0}$9.7 & 17.8  & OFHF \# 231 \nl
$\phantom{00}$2 &  8 30 36.0 & 65 52 36.9 &  56477 & 117 & $\phantom{0}$3.7 & 17.5  & OFHF \# 235 \nl
$\phantom{00}$3 &  8 30 37.1 & 65 49 27.8 &  56895 & $\phantom{0}$98 &  $\phantom{0}$7.4 & 19.2  &  \nl
$\phantom{00}$4 &  8 30 37.7 & 65 50 58.9 &  55001 & 113 &  $\phantom{0}$3.9 & 19.8  &  \nl
$\phantom{00}$5 &  8 30 38.1 & 65 51 39.8 &  57564 & 159 &  $\phantom{0}$6.5 & 18.1  &  \nl
$\phantom{00}$6 &  8 30 38.8 & 65 52 50.6 &  56874 & 112 &  $\phantom{0}$5.3 & 19.3  &  \nl
$\phantom{00}$7 &  8 30 39.1 & 65 49 49.0 &  54987 & 113 &  $\phantom{0}$5.0 & 20.1  &  \nl
$\phantom{00}$8 &  8 30 39.1 & 65 49 16.7 &  54000 & 101 &  $\phantom{0}$6.8 & 19.2  &  \nl
$\phantom{00}$9 &  8 30 40.2 & 65 50 25.3 &  56223 & 150 & \nodata & 18.7  & BLG \nl
$\phantom{0}$10 &  8 30 40.3 & 65 48 27.1 &  56400 & $\phantom{0}$93 &  12.7 & 18.1  &  \nl
$\phantom{0}$11 &  8 30 41.5 & 65 50 46.1 &  51186 & 150 & \nodata & 19.1  & BLG \nl
$\phantom{0}$12 &  8 30 42.4 & 65 50 34.6 &  51030 & 101 & $\phantom{0}$6.5 & 17.4  & OFHF \# 225 \nl
$\phantom{0}$13 &  8 30 43.9 & 65 51 28.7 &  76726 & 111 &  $\phantom{0}$4.3 & 18.9  & BG \nl
$\phantom{0}$14 &  8 30 46.3 & 65 51 12.4 &  70969 & $\phantom{0}$97 &  $\phantom{0}$4.7 & 20.3  & BG \nl
$\phantom{0}$15 &  8 30 48.2 & 65 52 33.1 &  54477 & 115 &  $\phantom{0}$8.2 & 18.6  &  \nl
$\phantom{0}$16 &  8 30 49.0 & 65 50 00.3 &  52381 & 108 &  $\phantom{0}$5.2 & 18.5  &  \nl
$\phantom{0}$17 &  8 30 49.7 & 65 48 52.2 &  55987 & 112 &  $\phantom{0}$4.4 & 20.1  &  \nl
$\phantom{0}$18 &  8 30 50.4 & 65 51 06.7 &  51728 & 118 &  $\phantom{0}$4.7 & 18.6  &  \nl
$\phantom{0}$19 &  8 30 51.4 & 65 48 38.7 &  53033 & 104 &  $\phantom{0}$4.9 & 19.1  &  \nl
$\phantom{0}$20 &  8 30 52.7 & 65 49 25.6 &  82649 & 109 &  $\phantom{0}$3.7 & 19.1  & BG  \nl
$\phantom{0}$21 &  8 30 52.8 & 65 48 48.8 &  57751 & 112 &  $\phantom{0}$4.7 & 18.3  &  \nl
$\phantom{0}$22 &  8 30 53.6 & 65 50 42.7 &  55998 & 103 &  $\phantom{0}$7.6 & 18.2  &  \nl
$\phantom{0}$23 &  8 30 53.9 & 65 51 23.1 &  56261 & $\phantom{0}$97 &  $\phantom{0}$8.6 & 18.3  &  \nl
$\phantom{0}$24 &  8 30 54.5 & 65 51 24.4 &  56023 & 102 &  $\phantom{0}$7.9 & 18.3  &  \nl
$\phantom{0}$25 &  8 30 54.9 & 65 52 16.8 &  53773 & 101 &  $\phantom{0}$9.0 & 17.7  &  \nl
$\phantom{0}$26 &  8 30 55.4 & 65 50 05.8 &  56745 & $\phantom{0}$91 &  11.8 & 18.6  &  \nl
$\phantom{0}$27 &  8 30 55.5 & 65 50 40.8 &  53850 & $\phantom{0}$98 &  11.4 & 19.0  &  \nl
$\phantom{0}$28 &  8 30 55.6 & 65 51 37.4 &  54583 & 101 &  $\phantom{0}$6.1 & 19.7  &  \nl
$\phantom{0}$29 &  8 30 55.8 & 65 49 00.7 &  56617 & 100 &  $\phantom{0}$7.3 & 18.5  &  \nl
$\phantom{0}$30 &  8 30 56.0 & 65 51 02.4 &  55729 & 112 &  $\phantom{0}$4.4 & 19.3  &  \nl
$\phantom{0}$31 &  8 30 56.2 & 65 49 46.4 &  56121 & $\phantom{0}$95 &  10.9 & 18.3  &  \nl
$\phantom{0}$32 &  8 30 56.2 & 65 49 11.4 &  55361 & $\phantom{0}$94 &  11.1 & 18.2  &  \nl
$\phantom{0}$33 &  8 30 56.7 & 65 50 57.7 &  57783 & $\phantom{0}$99 &  $\phantom{0}$6.7 & 19.8  &  \nl
$\phantom{0}$34 &  8 30 57.6 & 65 50 29.6 &  55015 & $\phantom{0}$90 & 16.7 & 17.2  & BCM, OFHF \# 201 \nl
$\phantom{0}$35 &  8 30 58.5 & 65 51 16.2 &  56502 & 106 &  $\phantom{0}$7.2 & 19.7  &  \nl
$\phantom{0}$36 &  8 30 58.7 & 65 48 47.7 &  55812 & 113 &  $\phantom{0}$5.1 & 20.0  &  \nl
$\phantom{0}$37 &  8 30 59.0 & 65 50 22.9 &  53306 & $\phantom{0}$94 &  12.8 & 17.5  &  \nl
$\phantom{0}$38 &  8 30 59.7 & 65 52 12.9 &  54722 & 102 &  $\phantom{0}$6.5 & 20.2  &  \nl
$\phantom{0}$39 &  8 31 00.6 & 65 50 22.0 &  54315 & 108 &  $\phantom{0}$4.5 & 20.7  &  \nl
$\phantom{0}$40 &  8 31 01.1 & 65 50 42.8 &  57303 & 107 &  $\phantom{0}$4.6 & 19.4  &  \nl
$\phantom{0}$41 &  8 31 02.3 & 65 49 36.6 &  54785 & 101 &  $\phantom{0}$7.8 & 18.9  &  \nl
$\phantom{0}$42 &  8 31 03.5 & 65 50 47.5 &  53476 & $\phantom{0}$93 &  14.1 & 17.4  &  OFHF \# 224 \nl
$\phantom{0}$43 &  8 31 05.4 & 65 51 04.1 &  50874 & $\phantom{0}$91 &  13.8 & 18.2  &  \nl
$\phantom{0}$44 &  8 31 05.6 & 65 48 46.5 &  44333 & 138 &  $\phantom{0}$4.9 & 18.0  & FG \nl
$\phantom{0}$45 &  8 31 07.9 & 65 49 24.2 &  54959 & $\phantom{0}$95 &  $\phantom{0}$8.9 & 18.8  &  \nl
$\phantom{0}$46 &  8 31 10.7 & 65 49 16.5 &  56906 & $\phantom{0}$98 &  $\phantom{0}$6.8 & 18.1  & OFHF \# 218 \nl
$\phantom{0}$47 &  8 31 11.1 & 65 51 46.8 &  53212 & $\phantom{0}$95 &  $\phantom{0}$5.9 & 19.9  &  \nl
$\phantom{0}$48 &  8 31 11.6 & 65 49 27.7 &  54243 & 106 &  $\phantom{0}$6.9 & 18.9  &  \nl
$\phantom{0}$49 &  8 31 12.3 & 65 52 42.4 &  53281 & 105 & $\phantom{0}$6.0 & 18.0  & OFHF \# 234 \nl
$\phantom{0}$50 &  8 31 13.2 & 65 49 35.4 &  56897 & $\phantom{0}$98 &  $\phantom{0}$6.8 & 18.1  &  \nl
$\phantom{0}$51 &  8 31 14.2 & 65 50 45.5 &  53624 & 123 &  $\phantom{0}$7.3 & 18.5  &  \nl
$\phantom{0}$52 &  8 31 16.3 & 65 51 36.2 &  55909 & 114 &  $\phantom{0}$5.3 & 19.4  &  \nl
$\phantom{0}$53 &  8 31 19.0 & 65 51 06.6 &  54356 & 129 &  $\phantom{0}$4.4 & 19.8  &  \nl
$\phantom{0}$54 &  8 31 22.5 & 65 49 13.1 &  54226 & 116 &  $\phantom{0}$6.0 & 18.6  &  \nl
$\phantom{0}$55 &  8 31 23.3 & 65 52 33.1 &  54307 & 114 &  $\phantom{0}$5.6 & 19.2  &  \nl
$\phantom{0}$56 &  8 30 35.8 & 65 52 32.3 &  \nodata  &   \nodata   &  \nodata  & 18.3  &  \nl
$\phantom{0}$57 &  8 30 36.1 & 65 54 07.7 &  \nodata  &   \nodata   &  \nodata  & 17.9  &  \nl
$\phantom{0}$58 &  8 30 36.1 & 65 49 50.7 &  \nodata  &   \nodata   &  \nodata  & 19.7  &  \nl
$\phantom{0}$58 &  8 30 37.4 & 65 52 22.8 &  \nodata  &   \nodata   &  \nodata  & 19.5  & o \nl
$\phantom{0}$60 &  8 30 38.4 & 65 53 33.0 &  \nodata  &   \nodata   &  \nodata  & 17.9  &  \nl
$\phantom{0}$61 &  8 30 39.4 & 65 50 35.1 &  \nodata  &   \nodata   &  \nodata  & 19.3  & o \nl
$\phantom{0}$62 &  8 30 41.7 & 65 52 17.6 &  \nodata  &   \nodata   &  \nodata  & 20.3  &  \nl
$\phantom{0}$63 &  8 30 42.1 & 65 49 07.9 &  \nodata  &   \nodata   &  \nodata  & 20.3  & o \nl
$\phantom{0}$64 &  8 30 43.0 & 65 49 45.6 &  \nodata  &   \nodata   &  \nodata  & 20.6  &  \nl
$\phantom{0}$65 &  8 30 43.1 & 65 50 18.4 &  \nodata  &   \nodata   &  \nodata  & 12.2  & gsc, o \nl 	
$\phantom{0}$66 &  8 30 43.5 & 65 50 43.8 &  \nodata  &   \nodata   &  \nodata  & 18.2  & OFHF \# 227, BG, o \nl
$\phantom{0}$67 &  8 30 43.9 & 65 48 01.3 &  \nodata  &   \nodata   &  \nodata  & 19.7  & o \nl
$\phantom{0}$68 &  8 30 44.0 & 65 50 34.3 &  \nodata  &   \nodata   &  \nodata  & 20.4  & o \nl
$\phantom{0}$69 &  8 30 44.3 & 65 50 40.8 &  \nodata  &   \nodata   &  \nodata  & 19.4  &  \nl
$\phantom{0}$70 &  8 30 44.6 & 65 52 12.5 &  \nodata  &   \nodata   &  \nodata  & 19.9  &  \nl
$\phantom{0}$71 &  8 30 44.9 & 65 51 55.1 &  \nodata  &   \nodata   &  \nodata  & 19.4  & o \nl
$\phantom{0}$72 &  8 30 45.4 & 65 48 25.1 &  \nodata  &   \nodata   &  \nodata  & 18.8  &  \nl
$\phantom{0}$73 &  8 30 45.4 & 65 50 42.7 &  \nodata  &   \nodata   &  \nodata  & 20.6  &  \nl
$\phantom{0}$74 &  8 30 46.4 & 65 52 54.7 &  \nodata  &   \nodata   &  \nodata  & 20.3  & o \nl
$\phantom{0}$75 &  8 30 46.7 & 65 49 47.6 &  \nodata  &   \nodata   &  \nodata  & 19.4  &  \nl
$\phantom{0}$76 &  8 30 47.4 & 65 50 59.9 &  \nodata  &   \nodata   &  \nodata  & 15.9  &  ps \nl
$\phantom{0}$77 &  8 30 47.7 & 65 49 27.4 &  \nodata  &   \nodata   &  \nodata  & 20.7  &  \nl
$\phantom{0}$78 &  8 30 48.3 & 65 49 57.9 &  \nodata  &   \nodata   &  \nodata  & 19.1  &  \nl
$\phantom{0}$79 &  8 30 49.2 & 65 50 50.6 &  \nodata  &   \nodata   &  \nodata  & 14.5  &  gsc \nl
$\phantom{0}$80 &  8 30 49.5 & 65 47 57.1 &  \nodata  &   \nodata   &  \nodata  & 19.8  & o \nl
$\phantom{0}$81 &  8 30 50.7 & 65 48 34.6 &  \nodata  &   \nodata   &  \nodata  & 20.0  & o \nl
$\phantom{0}$82 &  8 30 50.9 & 65 51 23.1 &  \nodata  &   \nodata   &  \nodata  & 15.9  &  ps \nl
$\phantom{0}$83 &  8 30 51.0 & 65 49 28.2 &  \nodata  &   \nodata   &  \nodata  & 18.0  & s,o \nl
$\phantom{0}$84 &  8 30 51.3 & 65 51 12.1 &  \nodata  &   \nodata   &  \nodata  & 18.3  &  \nl
$\phantom{0}$85 &  8 30 51.3 & 65 51 30.4 &  \nodata  &   \nodata   &  \nodata  & 19.8  & o \nl
$\phantom{0}$86 &  8 30 51.9 & 65 52 25.7 &  \nodata  &   \nodata   &  \nodata  & 18.0  & o \nl
$\phantom{0}$87 &  8 30 52.2 & 65 49 20.4 &  \nodata  &   \nodata   &  \nodata  & 20.2  &  \nl
$\phantom{0}$88 &  8 30 52.4 & 65 51 03.2 &  \nodata  &   \nodata   &  \nodata  & 20.2  & o \nl
$\phantom{0}$89 &  8 30 52.5 & 65 48 58.7 &  \nodata  &   \nodata   &  \nodata  & 20.3  &  \nl
$\phantom{0}$90 &  8 30 52.7 & 65 52 21.3 &  \nodata  &   \nodata   &  \nodata  & 19.4  & o \nl
$\phantom{0}$91 &  8 30 54.7 & 65 50 22.8 &  \nodata  &   \nodata   &  \nodata  & 20.0  &  \nl
$\phantom{0}$92 &  8 30 54.7 & 65 49 24.1 &  \nodata  &   \nodata   &  \nodata  & 19.5  & o \nl
$\phantom{0}$93 &  8 30 55.0 & 65 52 08.4 &  \nodata  &   \nodata   &  \nodata  & 18.6  &  \nl
$\phantom{0}$94 &  8 30 55.8 & 65 50 49.4 &  \nodata  &   \nodata   &  \nodata  & 20.6  &  \nl
$\phantom{0}$95 &  8 30 56.0 & 65 52 04.6 &  \nodata  &   \nodata   &  \nodata  & 18.6  &  \nl
$\phantom{0}$96 &  8 30 57.0 & 65 49 48.6 &  \nodata  &   \nodata   &  \nodata  & 20.1  & o \nl
$\phantom{0}$97 &  8 30 57.8 & 65 50 17.3 &  \nodata  &   \nodata   &  \nodata  & 19.3  &  \nl
$\phantom{0}$98 &  8 30 58.2 & 65 52 26.5 &  \nodata  &   \nodata   &  \nodata  & 20.0  &  \nl
$\phantom{0}$99 &  8 30 58.4 & 65 50 17.2 &  \nodata  &   \nodata   &  \nodata  & 20.2  &  \nl
100 &  8 30 58.7 & 65 50 07.0 &  \nodata  &   \nodata   &  \nodata  & 20.5  &  \nl
101 &  8 30 58.7 & 65 48 02.3 &  \nodata  &   \nodata   &  \nodata  & 19.6  &  \nl
102 &  8 30 58.9 & 65 50 42.6 &  \nodata  &   \nodata   &  \nodata  & 19.7  &  \nl
103 &  8 30 59.2 & 65 50 52.2 &  \nodata  &   \nodata   &  \nodata  & 18.1  & ps, o \nl
104 &  8 30 59.5 & 65 50 20.6 &  \nodata  &   \nodata   &  \nodata  & 17.8  &  \nl
105 &  8 30 59.6 & 65 50 57.2 &  \nodata  &   \nodata   &  \nodata  & 19.5  & o\nl
106 &  8 30 59.7 & 65 49 34.6 &  \nodata  &   \nodata   &  \nodata  & 20.5  &  \nl
107 &  8 30 59.9 & 65 50 47.8 &  \nodata  &   \nodata   &  \nodata  & 20.0  & o \nl
108 &  8 31 00.3 & 65 51 47.9 &  \nodata  &   \nodata   &  \nodata  & 14.9  &  gsc \nl
109 &  8 31 00.6 & 65 48 59.2 &  \nodata  &   \nodata   &  \nodata  & 19.4  & o \nl
110 &  8 31 00.6 & 65 51 37.6 &  \nodata  &   \nodata   &  \nodata  & 19.6  &  \nl
111 &  8 31 00.8 & 65 48 41.3 &  \nodata  &   \nodata   &  \nodata  & 20.0  & o \nl
112 &  8 31 01.5 & 65 48 08.7 &  \nodata  &   \nodata   &  \nodata  & 20.2  &  \nl
113 &  8 31 03.1 & 65 50 02.4 &  \nodata  &   \nodata   &  \nodata  & 19.9  &  \nl
114 &  8 31 03.2 & 65 49 55.8 &  \nodata  &   \nodata   &  \nodata  & 17.6  &  \nl
115 &  8 31 03.4 & 65 48 51.7 &  \nodata  &   \nodata   &  \nodata  & 20.1  &  \nl
116 &  8 31 04.1 & 65 50 09.1 &  \nodata  &   \nodata   &  \nodata  & 19.0  &  \nl
117 &  8 31 05.2 & 65 47 52.9 &  \nodata  &   \nodata   &  \nodata  & 19.4  & o \nl
118 &  8 31 05.6 & 65 51 56.5 &  \nodata  &   \nodata   &  \nodata  & 20.7  &  \nl
119 &  8 31 05.6 & 65 51 08.9 &  \nodata  &   \nodata   &  \nodata  & 19.2  & o \nl
120 &  8 31 05.6 & 65 50 43.6 &  \nodata  &   \nodata   &  \nodata  & 19.9  & pb, o \nl
121 &  8 31 06.4 & 65 48 14.7 &  \nodata  &   \nodata   &  \nodata  & 19.9  &  \nl
122 &  8 31 06.6 & 65 49 38.4 &  \nodata  &   \nodata   &  \nodata  & 20.1  &  \nl
123 &  8 31 06.9 & 65 50 17.8 &  \nodata  &   \nodata   &  \nodata  & 16.9  &  ps \nl
124 &  8 31 07.2 & 65 49 19.4 &  \nodata  &   \nodata   &  \nodata  & 18.4  & o \nl
125 &  8 31 07.3 & 65 51 35.1 &  \nodata  &   \nodata   &  \nodata  & 20.4  &  \nl
126 &  8 31 07.9 & 65 50 43.0 &  \nodata  &   \nodata   &  \nodata  & 19.5  &  \nl
127 &  8 31 07.9 & 65 50 05.7 &  \nodata  &   \nodata   &  \nodata  & 20.0  & o \nl
128 &  8 31 08.0 & 65 49 35.7 &  \nodata  &   \nodata   &  \nodata  & 19.4  & o \nl
129 &  8 31 08.3 & 65 51 43.5 &  \nodata  &   \nodata   &  \nodata  & 19.3  & o \nl
130 &  8 31 09.9 & 65 48 40.6 &  \nodata  &   \nodata   &  \nodata  & 19.8  &  \nl
131 &  8 31 10.1 & 65 48 11.9 &  \nodata  &   \nodata   &  \nodata  & 17.0  & ps \nl
132 &  8 31 10.4 & 65 47 48.3 &  \nodata  &   \nodata   &  \nodata  & 20.2  &  \nl
133 &  8 31 11.0 & 65 51 41.3 &  \nodata  &   \nodata   &  \nodata  & 19.8  & o \nl
134 &  8 31 11.2 & 65 49 52.5 &  \nodata  &   \nodata   &  \nodata  & 20.6  &  \nl
135 &  8 31 13.8 & 65 49 36.8 &  \nodata  &   \nodata   &  \nodata  & 18.1  & o \nl
136 &  8 31 16.2 & 65 49 42.0 &  \nodata  &   \nodata   &  \nodata  & 20.8  &  \nl
137 &  8 31 16.2 & 65 49 03.7 &  \nodata  &   \nodata   &  \nodata  & 17.9  &  \nl
138 &  8 31 16.4 & 65 50 26.9 &  \nodata  &   \nodata   &  \nodata  & 18.1  &  \nl
139 &  8 31 18.6 & 65 52 10.3 &  \nodata  &   \nodata   &  \nodata  & 18.1  &  \nl
140 &  8 31 18.6 & 65 48 38.4 &  \nodata  &   \nodata   &  \nodata  & 20.0  &  \nl
141 &  8 31 19.1 & 65 50 01.6 &  \nodata  &   \nodata   &  \nodata  & 19.2  &  \nl
142 &  8 31 19.3 & 65 52 51.4 &  \nodata  &   \nodata   &  \nodata  & 20.0  & o \nl
143 &  8 31 19.5 & 65 49 28.3 &  \nodata  &   \nodata   &  \nodata  & 19.5  & o \nl
144 &  8 31 20.8 & 65 49 47.6 &  \nodata  &   \nodata   &  \nodata  & 20.1  &  \nl
145 &  8 31 22.6 & 65 51 31.7 &  \nodata  &   \nodata   &  \nodata  & 18.0  &  \nl
146 &  8 31 23.4 & 65 48 19.4 &  \nodata  &   \nodata   &  \nodata  & 20.6  &  \nl
147 &  8 31 23.7 & 65 52 20.9 &  \nodata  &   \nodata   &  \nodata  & 20.3  &  \nl
\enddata
\tablenotetext{}{NOTES}
\tablenotetext{}{BCM: brightest cluster member}
\tablenotetext{}{OFHF: galaxy also in Oegerle et al.~1991 sample}
\tablenotetext{}{BLG: galaxy with strong Balmer absorption lines}
\tablenotetext{}{BG: confirmed background galaxy}
\tablenotetext{}{FG: confirmed foreground galaxy}
\tablenotetext{}{o: object was observed spectroscopically but the crosscorrelation gave a low TDR value.}
\tablenotetext{}{gsc: Guide Star Catalog Star. The magnitude for \# 65 is from the GSC.}
\tablenotetext{}{pb: We determine that this object is a possible backgroung galaxy. Its crosscorrelation yielded a velocity $\sim$ 77,000 km s$^{-1}$ with a low TDR value of 3.4.}
\tablenotetext{}{s: stellar spectrum}
\tablenotetext{}{ps: object brighter than the BCM}
\end{deluxetable}

\begin{deluxetable}{c c c c}
\tablenum{3}
\tablewidth{0pt}
\tablecaption{Velocity Data}
\tablehead{
\colhead{} & \colhead{Number of} & \colhead{} & 
   \colhead{$\sigma_{\rm LOS}$}\\
\colhead{Sample} & \colhead{Galaxies}& \colhead{Redshift} 
  & \colhead{(km s$^{-1}$)}}
\startdata
All Galaxies  & 77 &  0.18285$^{+0.00045}_{-0.00064}$
   & 1390$^{+120}_{-110}$ \nl & & \nl
Our Sample    & 51 &  0.18373$^{+0.00067}_{-0.00102}$
   & 1500$^{+190}_{-110}$ \nl & & \nl
OHFH          & 33 &  0.18170$^{+0.00094}_{-0.00074}$        
   & 1230$^{+250}_{-140}$ \nl & & \nl
Central Galaxies ($R < 4\farcm5$)   & 54 & 0.18347$^{+0.00082}_{-0.00084}$
   & 1430$^{+170}_{-140}$ \nl & & \nl
Outer Galaxies ($R > 4\farcm5$)   & 23 & 0.18110$^{+0.00072}_{-0.00064}$
   & 1050$^{+270}_{-190}$ \nl & & \nl
\enddata
\end{deluxetable}

\clearpage
\begin{figure}
\plotfiddle{f1.ps}{4in}{270}{100}{100}{-400}{450}
\end{figure}

\clearpage
\begin{figure}
\plotfiddle{f2.ps}{4in}{0}{100}{100}{-300}{-300}
\end{figure}

\clearpage
\begin{figure}
\plotfiddle{f3.ps}{4in}{270}{100}{100}{-400}{450}
\end{figure}

\clearpage
\begin{figure}
\plotfiddle{f4.ps}{4in}{0}{100}{100}{-300}{-200}
\end{figure}

\clearpage
\begin{figure}
\plotfiddle{f5.ps}{4in}{0}{100}{100}{-300}{-300}
\end{figure}

\clearpage
\begin{figure}
\plotfiddle{f6.ps}{4in}{0}{100}{100}{-300}{-300}
\end{figure}

\clearpage
\begin{figure}
\plotfiddle{f7.ps}{4in}{0}{100}{100}{-300}{-300}
\end{figure}

\end{document}